\documentstyle[twocolumn,aps,epsfig]{revtex}

\begin{document}

\title{Inhomogeneous relaxation in vibrated granular media: consolidation
 waves}

\author{G C Barker}
\address{
Institute of Food Research,\\
Norwich Research Park,\\
Colney Lane, Norwich NR4 7UA, UK\\
email: barker@bbsrc.ac.uk}
\author{Anita Mehta}
\address{
S N Bose National Centre for Basic Sciences\\
Block JD Sector III Salt Lake\\
Calcutta 700091, India\\
email: anita@boson.bose.res.in}

\date{\today}
\maketitle

\abstract{We investigate the spatial dependence of the density of
vibrated granular beds, using simulations based on a hybrid Monte Carlo
algorithm. We find that the initial consolidation is
typically inhomogeneous, both in the presence of a constant
shaking intensity and when the granular bed is submitted to
`annealed shaking'. We also present a theoretical model which
explains such inhomogeneous relaxation in terms of a `consolidation
wave', in good qualitative agreement with our simulations. Our
results are also in qualitative agreement with recent experiments.}

\section{Introduction}
In recent years the issue of compaction in granular media has
attracted experimental (\cite{knight},\cite{nowak},
\cite{nowak1}) as well as theoretical (\cite{dg}, \cite{dg1}, \cite{sam}
\cite{PRL}, \cite{PRAE}, \cite{ant}), interest in the physics
community. In particular the important issue of compactivity
\cite{sam1}, as an effective temperature for powders in the
quasi-static regime, has intrigued researchers \cite{nowak1} who
have sought, as a consequence, to investigate further the
implications of such a definition.

While this effective temperature was originally defined in terms
of thermodynamic analogies for a {\it static} powder, some
inconsistencies \cite{sam1} have motivated work on other possible
definitions. A Langevin dynamics approach, used to model vibrated
powders \cite{ars}, was able to introduce the compactivity as the
natural interpolation of the conventional `granular temperature'
\cite{savage} as the intensity of vibration was decreased; this
led to its interpretation as the effective temperature
corresponding to the slow dynamical modes of the system. This
interpretation of the compactivity was recently reformulated at a
more microscopic level, in terms of a p-spin model of granular
dynamics \cite{jorge}.

As the compactivity parameter is closely related to density
variations \cite{ambook}, recent experimental investigations have
focused on spatiotemporal density fluctuations in a shaken powder
\cite{knight}. Recent molecular dynamics simulations
\cite{swinney} have also highlighted this issue by focusing on
the necessarily inhomogeneous density fluctuations which arise
when a granular system is forced by the motion of its boundaries.
In this paper, we examine two aspects of spatial inhomogeneities:
\begin{itemize}
\item First, we examine spatial  density fluctuations in the presence
of a constant shaking intensity, and find definite evidence of
inhomogeneous relaxation at least at short times. Additionally,
as in experiment, \cite{knight}, our simulations suggest that the
degree of inhomogeneity is {\it less}  at higher intensities of
shaking. We put our observations into perspective by suggesting
that the density of a vibrated granular bed initially relaxes by
the propagation of a `consolidation wave'; this behaviour can be
represented by a theoretical model that is in good qualitative
agreement with our simulations.
\item Second, we examine the response of the granular bed
to an `annealed shaking' procedure \cite{nowak}; the
consolidation along the socalled reversible branch of the
compaction curve shows evidence of strong spatial
inhomogeneities, which we demonstrate via the configurational
histories of the particles. Compaction proceeds from the walls,
and propagates like a consolidation wave; this is reminiscent of
earlier studies of transitions to crystallinity \cite{jpcm} in
shaken granular beds, and has also been observed in a recent
experiment \cite{rods} on the appearance of `nematic' ordering in
an assembly of rods.
\end{itemize}

\section{Procedure; the hybrid Monte Carlo model}
Our simulations use uniform hard spheres, subjected to
non-sequential reorganizations which represent the effect of
shaking. A  variable  shaking  amplitude $A$ is parametrised in
units of the particle diameter; thus for example, $A=1.0$ means
that shaken particles are able to move longitudinally and
laterally by, $~1$ particle diameter (subject to volume
exclusions) during a shake cycle. The details of the shaking
algorithm have been  discussed elsewhere
(\cite{PRL},\cite{ambook}). Briefly, one cycle of vibration of
the granular assembly (corresponding to one timestep of our
simulation) is modelled by:
\begin{enumerate}
\item a vertical dilation of the granular bed, in proportion
to the shaking amplitude $A$
\item a stochastic
rearrangement of the individual particles in an external field and
with available free volume proportional to the shaking amplitude
\item and finally a {\it cooperative} recompression of the assembly
as each grain lands on the substrate alone or with neighbours (in
the latter case, arches could form).
\end{enumerate}

This algorithm has both static and dynamic disorder. The former
is incorporated via the non-lattice-based nature of the
simulations, while the latter is a consequence of the stochastic
step, which allows particles the freedom to move in planes
perpendicular to the direction of vibration, in proportion to the
driving force. Collective structures, such as arches, are also
allowed to form and propagate, because of the inclusion of the
cooperative recompression step. In the absence of this step,
particles would land sequentially, and the configurations would
include only simple three-point stability conditions.

\section{Results on spatial dependence of packing fraction}
In this section, we explore first numerically and then
analytically, the spatial dependence of the packing fraction
$\phi$ of a shaken powder. We were guided in our investigations
by the knowledge that typically, experiments rely on {\it
indirect} knowledge of spatial variations of material properties,
since the use of 3d techniques, such as MRI \cite{eiichi}, to make
quantitative measurements, is still not very widespread.
Consequently, we look at macroscopic quantities such as the
average packing fraction $<\phi>$ and the mean particle height
$Z(t)$, and extract information
on spatial inhomogeneities of $<\phi>$, indirectly .
 In the analytical section we
discuss related issues, once again from a macroscopic point of
view.

\subsection{Numerical simulations}
\begin{figure}[ht]
\centerline{\epsfig{file=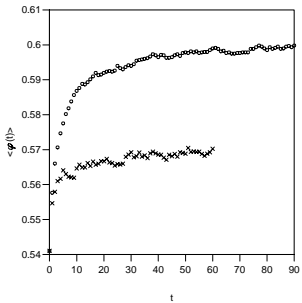,height=9cm,angle=270}}
\caption{Plot of mean  volume fraction $<\phi(t)>$ against time
$t$ for two different shaking amplitudes $A$. The open circles
correspond to $A=0.05$, while the crosses correspond to
$A=0.5$.}\label{Fig1}
\end{figure}

In Fig. \ref{Fig1}, we have plotted the mean packing fraction
$<\phi(t)>$ against time $t$ for two different amplitudes $A$,
where the angular brackets indicate an ensemble average over
distinct realizations (independent sequences of pseudo-random
numbers) of the noise.

The mean volume fraction, $<\phi(t)>$, is evaluated over the
central $80$ per cent of the vertical extent of the particle
assembly. This central zone is defined by two planes, with heights
$h_0=0.2Z, h_1=1.8Z$, placed symmetrically about the (volume
weighted)  mean particle height $Z(t)$. The granular medium is in
a box with a hard base, which has periodic boundaries in the
lateral directions. While all our simulations are initiated from
stable loose packings with $\phi \approx 0.54$, we note that the
rate of structural relaxation increases with the shaking
amplitude; the {\it time required to reach the steady state
decreases with increasing $A$}. More importantly however, we note
that the bulk of the relaxation to the higher packing fractions
occurs for the lower shaking amplitude, which is a direct
consequence \cite{PRL} of the predominance of  the slow dynamics
of cooperative relaxation. In this dynamical regime, collective
structures remain long-standing (their autocorrelation functions
decay only very slowly), and it is possible to show via the use of
displacement-displacement correlation functions \cite{PRAE} that
particles move together in {\it dynamical clusters}; the enhanced
compaction occurs when the net void space trapped in these
clusters is shaved off.
\begin{figure}[ht]
\centerline{\epsfig{file=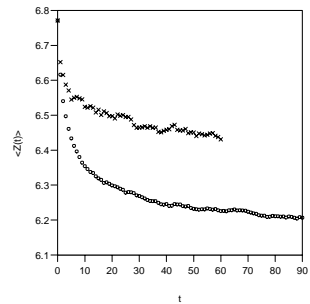,height=9cm,angle=270}}
\caption{Plot of mean  particle height $Z(t)$ against time $t$
for two different shaking amplitudes $A$. The open circles
correspond to $A=0.05$, while the crosses correspond to
$A=0.5$.}\label{Fig2}
\end{figure}

In Fig. \ref{Fig2}, we plot the mean particle height $Z(t)$
against time $t$; this is an important indicator of consolidation
as it shows the average distance of the particles from the base.
We notice that {\it the final value} of $<Z(t)>$ is smaller for
smaller shaking amplitudes. In other words, the particles are more
efficiently settled or `sedimented' for gentler shaking, and
consolidation is thus stronger.

\begin{figure}[ht]
\centerline{\epsfig{file=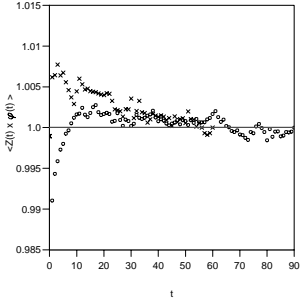,height=9cm,angle=270}}
\caption{Plot of mean (normalised) product of particle height
$<Z(t)\phi(t)>/<Z(\infty)\phi(\infty)>$ against time $t$ for two
different shaking amplitudes $A$. The open circles correspond to
$A=0.05$, while the crosses correspond to $A=0.5$.}\label{Fig3}
\end{figure}

These two indicators are independent measures of consolidation,
but reveal nothing about its nature. Their product
$<Z(t)\phi(t)>$, however, provides an insight into whether the
consolidation is homogeneous or inhomogeneous; this is plotted in
Figure \ref{Fig3}.

In the case  of {\it homogeneous} consolidation, the volume
fraction and mean particle height are related by a geometrical
(centroid) condition such that $\phi(t) Z(t)=$ constant. However
our plots of $<\phi(t) Z(t)>$ (normalised to the value at
$t=\infty$) (Fig. \ref{Fig3}) show systematic  deviations from
unity at early times and therefore  indicate  inhomogeneous
consolidation at these times.

The details of our findings can be explained by the results of a simple theory,
which we develop in the next subsection.

\subsection{Theory: consolidation waves}
Consider the simplest form of inhomogeneous consolidation, i.e.
the volume fraction changes discontinuously from an initial
uniform density, $\phi_0$, to a maximum density, $\phi_{max}$, at
each layer of a (three-dimensional) vibrated granular bed,
following the passage of a `consolidation wave'. We visualise the
process of consolidation to be analogous to that which is known
to occur in sedimentation \cite{gary}, when particles settle onto
a sediment at the bottom of a particulate suspension. In this way,
we imagine that shaken grains settle to the bottom of the vibrated
bed in a more compacted state than they were previously, and
that this settling defines the position of the front of the
consolidation wave.

Let $z_{max} (t)$ be the position of the highest layer with
density $\phi_{max}$ at time $t$; this then defines the position
of the front of the consolidation wave. Let
$z_0(t)$ be the position of the upper surface (characterised
by a density $\phi_0(t)$  at time $t$.
Assuming that consolidation is uniform in the plane perpendicular
to the direction of vibration (which we choose to be the
$z$-direction) , while it is inhomogeneous in the direction
of vibration, conservation of mass gives\\
\begin{eqnarray}
     \phi_{max}(t) z_{max}(t) +
      \phi_0(t) (z_0(t)-z_{max}(t)) & \nonumber\\
                                   =& \phi_0(0) z_0(0)
\end{eqnarray}\\
The position $Z(t)$ corresponding to the mean particle height  at time $t$ is:\\
\begin{eqnarray}
 Z(t) &  = & [(\phi_{max}(t)-\phi_0(t))z_{max}(t)^2 \\
 &            + & \phi_0(t)z_0(t)^2]/2\phi_0(0)z_0(0) \nonumber
\end{eqnarray}\\
The position of the consolidation wavefront $z_{max} (t)$
relative to the central zone, i.e. the region between planes
$h_0(t) = 0.2Z(t)$ and $h_1(t) = 0.8Z(t)$, defines the measured
mean volume fraction $<\phi (t)>$ at time $t$:\\
\begin{eqnarray}
<\phi(t)> & = & \phi(0)   \qquad \qquad \qquad \qquad       z_{max}(t)< h_0(t)\\
<\phi(t)> & = & [(\phi_{max}-\phi_0)z_{max}(t)  \nonumber\\
         &+& (1.8\phi(0)-0.2\phi_{max})Z(t))]/1.6Z(t), \nonumber\\
            && \qquad   \qquad \qquad  9h_0(t) > z_{max}(t) > h_0(t)\\
<\phi(t)>  &=& \phi_{max}       \qquad \qquad \qquad \qquad
z_{max}(t) > 9h_0(t)
\end{eqnarray}\\
Equations (2) and (3-5) are in qualitative agreement with Figures
\ref{Fig1} and \ref{Fig2}.

In order to explain the non-trivial behaviour of the product
$<\phi(t) Z(t)>$, we note that at large times, $t\to\infty$, we  must have:\\
\begin{eqnarray}
z_{max}(t) & = & z_0(t)=2Z(t)\\
\phi_{max}Z(t) & = & \phi(0)Z(0)
\end{eqnarray}\\
since the consolidation wavefront will have reached the upper
surface at large times. Then, using the above equations, and
simplifying, we find that for short times, the normalized product
$<\phi(t)Z(t)>$ (which approaches one as $t\to\infty$),
is given by\\
\begin{eqnarray}
[\phi(t)Z(t)]/[\phi_{max}Z(\infty)] & = & Z(t)/Z(0) \nonumber\\
                       & = & z_0(t)/z_0(0) + \nonumber\\
((\phi(0)/(\phi_{max}  -  \phi_0))(& [ &(z_0(0)-z_0(t))/z_0(0)]^2)
\end{eqnarray}\\

The first term on the right-hand-side is
expected to be less than one in the case of consolidation,
since the height of the upper surface
should decrease with time as the granular bed compacts. The
second term, in this case, is also clearly positive; its size, however,
depends on the size of the coefficient
$\phi(0)/(\phi_{max}-\phi(0))$. We have seen in Figure 1 that
small values of $\phi_{max}-\phi(0)$ correspond to large shaking
amplitudes $A$, and conversely.  This results implies, in agreement
with earlier work on compaction \cite{PRL}, that the overall
consolidation is larger for lower shaking amplitudes. Thus,  for large
$A$ the normalised product $<\phi(t)Z(t)>$ can be
greater than one ; for small shaking amplitudes, on the other
hand, the  value of the product can be less than one (the value
of $\phi(0)/(\phi_{max}-\phi(0))$ is small). These results
obtained in the short-time limit of this very simple model, are
in excellent agreement with the results of the simulation (Figure
\ref{Fig3}).

\section{Irreversible and reversible branches in
annealed shaking: more consolidation waves}
Recent explorations of
compaction in a shaken powder have involved the analogue of
'annealed cooling' \cite{nowak}; a granular bed, prepared at very
low density, is subjected to $t_{tap}$ taps at intensity $\Gamma$; the
intensity is then incremented in steps of $\delta\Gamma$. This results in
a compaction curve with two branches; an irreversible branch,
where increasing the tapping intensity leads to increasing
packing fractions, and a (so-called) reversible branch, where the
reverse is the case. We have simulated the `annealed shaking' of
a granular bed of $1300$ particles for $1425$ shakes, such that
$\delta\Gamma = 1/500$ and $t_{tap}=1$, corresponding to a
continuous variation of shaking intensity. As before, there is a
hard base in the lower $z$ direction, while there are periodic
boundary conditions in the $x$ and $y$ directions. The intensity
is increased from $A=0.05$ to $A=1.0$, decreased to $A=0.05$ and
increased back again to $A=1.0$.

\begin{figure}[ht]
\centerline{\epsfig{file=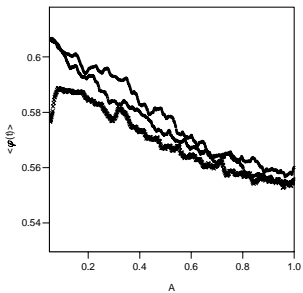,height=9cm,angle=270}}
\caption{Plot of packing fraction $\phi$ vs shaking intensity $A$
for the annealed shaking of a granular bed. Note the hysteresis
loop in the 'reversible branch', showing a fundamental
irreversibility.} \label{Fig4}
\end{figure}

 Our results  (Fig. \ref{Fig4}) show
that
 the packing fraction first
increases from $0.55$ along the irreversible branch before
decreasing at high values of $A$; it then increases as $A$ is
reduced along the socalled reversible branch. Any further changes
of shaking intensity cause it to remain on this branch, 
{\it except} for a small hysteresis loop; the presence of this
hysteresis loop indicates that even the socalled reversible
branch is {\em irreversible} in reality. We expect that the size
of this hysteresis loop would increase for large $\delta\Gamma$ and small
 $t_{tap}$, and conversely; the irreversibility of the `reversible'
branch consequently mirrors the degree of irreversibility of the
 annealed shaking process.

In order to look more microscopically at what occurs along
both branches of the compaction curve, we present
in Fig. 5 snapshots of the configurations of the spheres. Each
snapshot
corresponds to a distinct
 value of the shake intensity $A$ as the compaction
curves of Fig. \ref{Fig4} are swept.  The first three
configurations correspond to  $A$  {\it increasing} on
the irreversible branch, the second three, to $A$ {\it decreasing}
on the {\it reversible}
branch, and  the last three to $A$ {\it increasing} along the {\it reversible}
branch. We note, especially along the `reversible' trajectories
that a `crystalline'-like ordering appears to spread from the base
as the shake intensity is continuously decreased, which has a
wavelike
character.
The presence of this `consolidation wave'
 mirrors a recent experimental result, where ordering in an
assembly of rods
is seen to  originate at the walls and to
propagates like a wavefront \cite{rods}.

\begin{figure}
\begin{center}
\label{Fig5} \epsfxsize=2.5truecm\epsfysize=4truecm
{\epsfbox{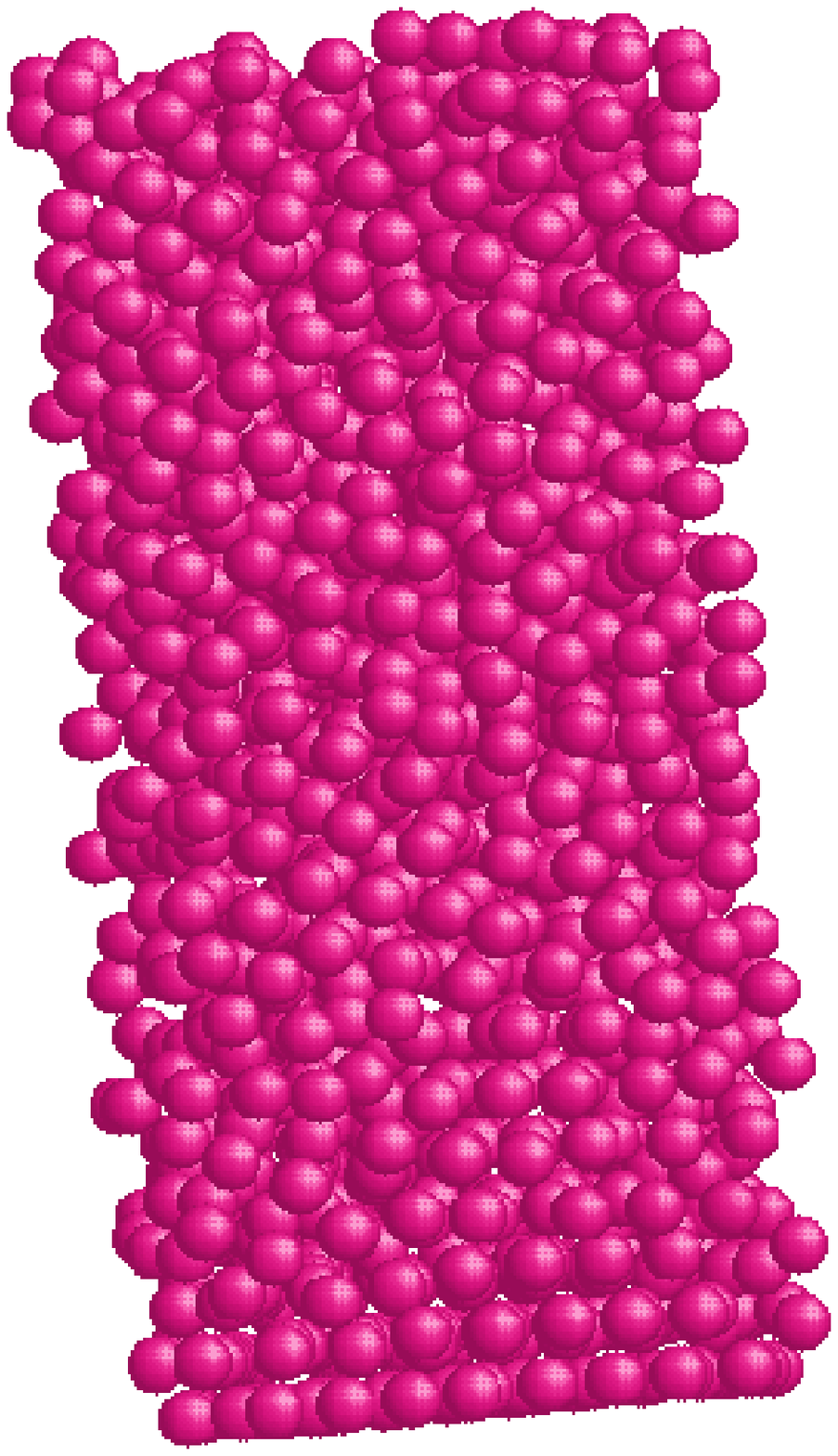}} \epsfxsize=2.5truecm\epsfysize=4truecm
{\epsfbox{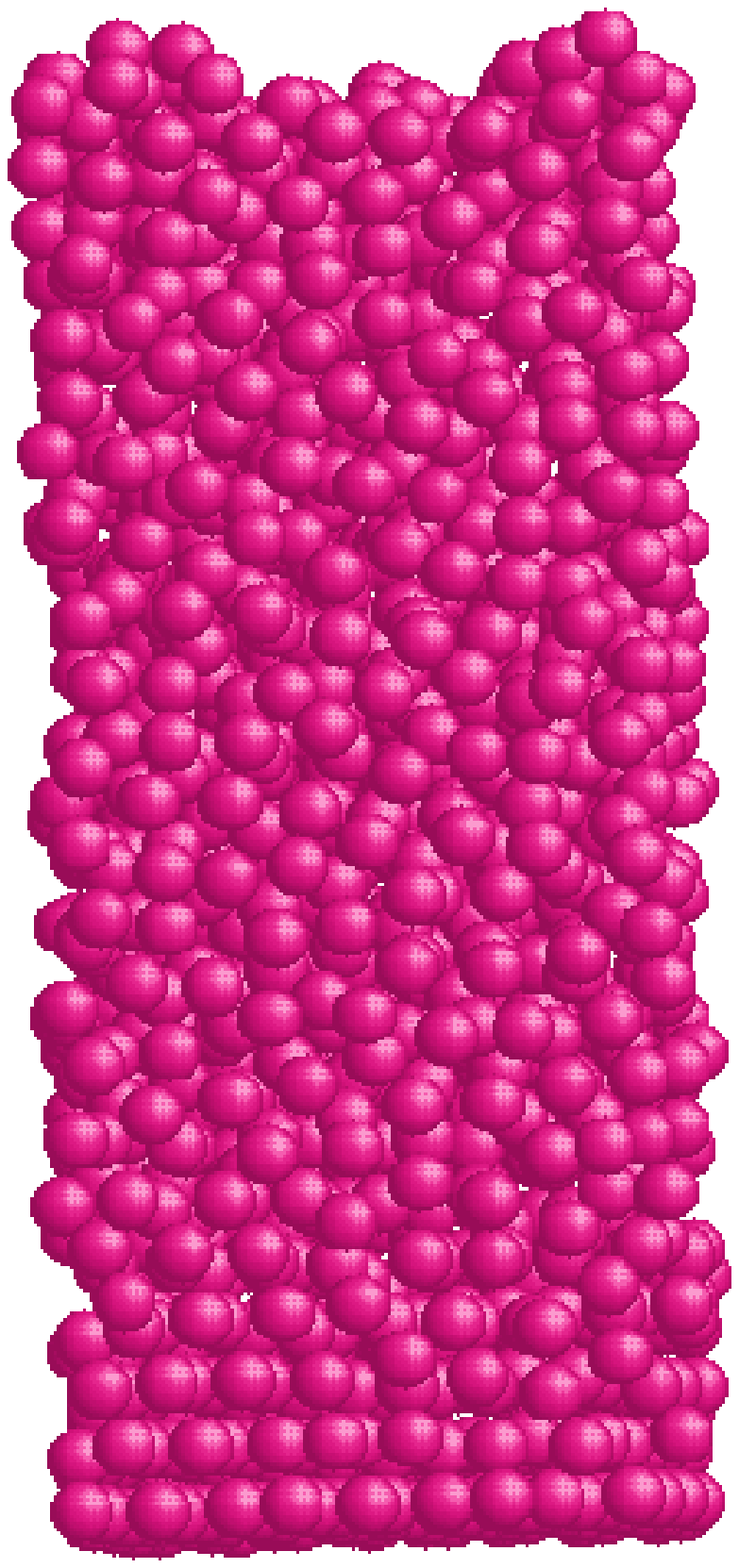}} \epsfxsize=2.5truecm\epsfysize=4truecm
{\epsfbox{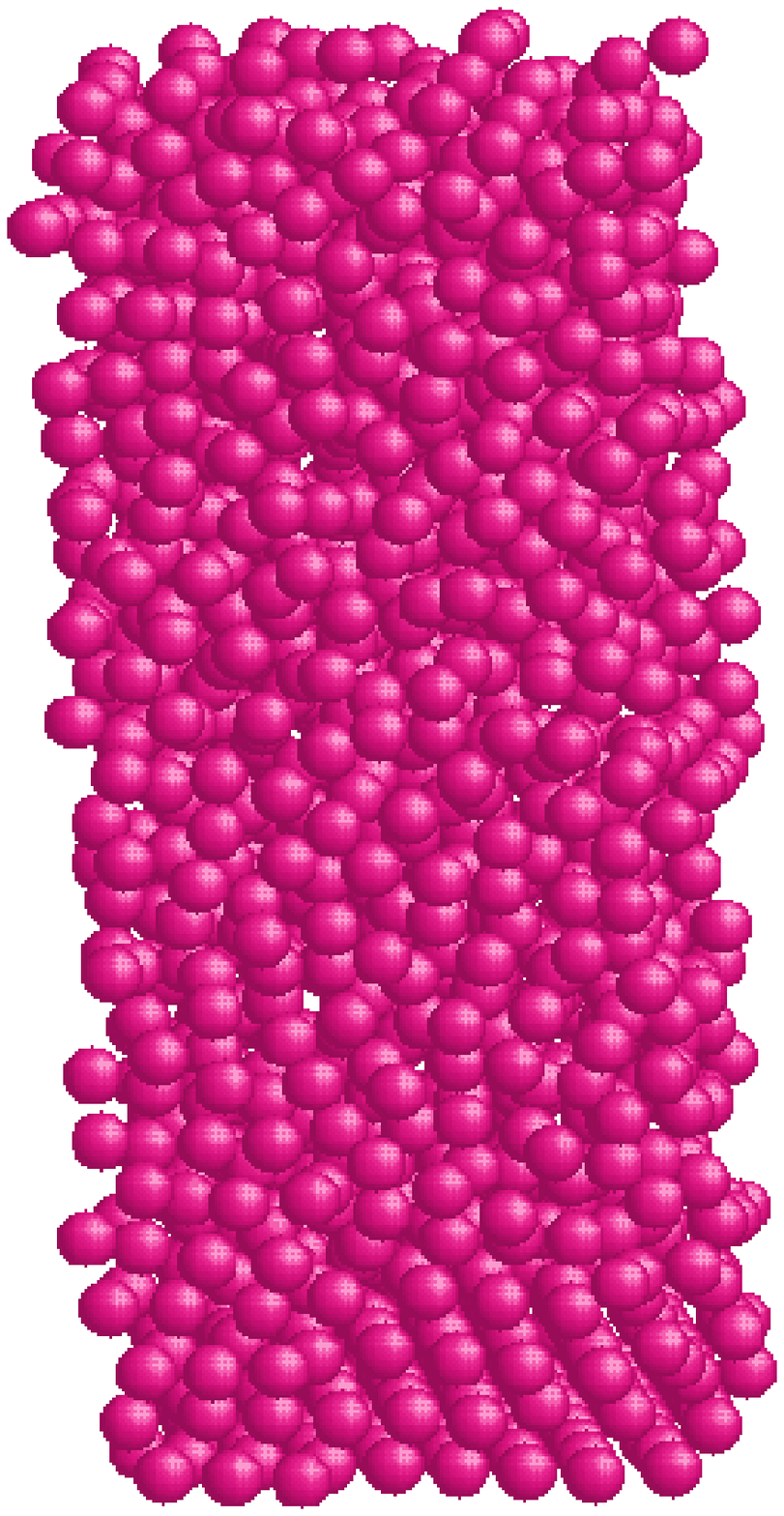}} \epsfxsize=2.5truecm\epsfysize=4truecm
{\epsfbox{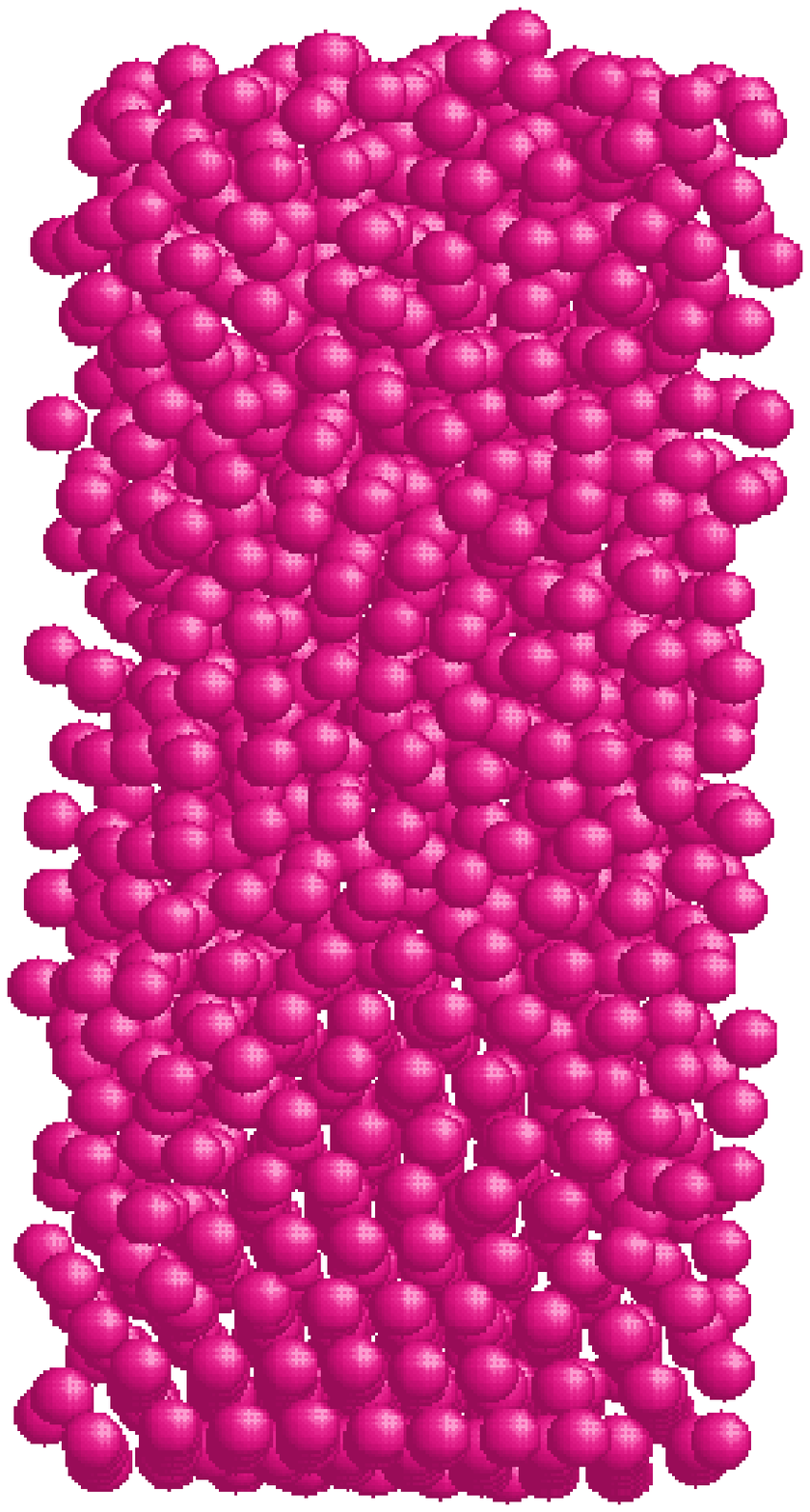}} \epsfxsize=2.5truecm\epsfysize=4truecm
{\epsfbox{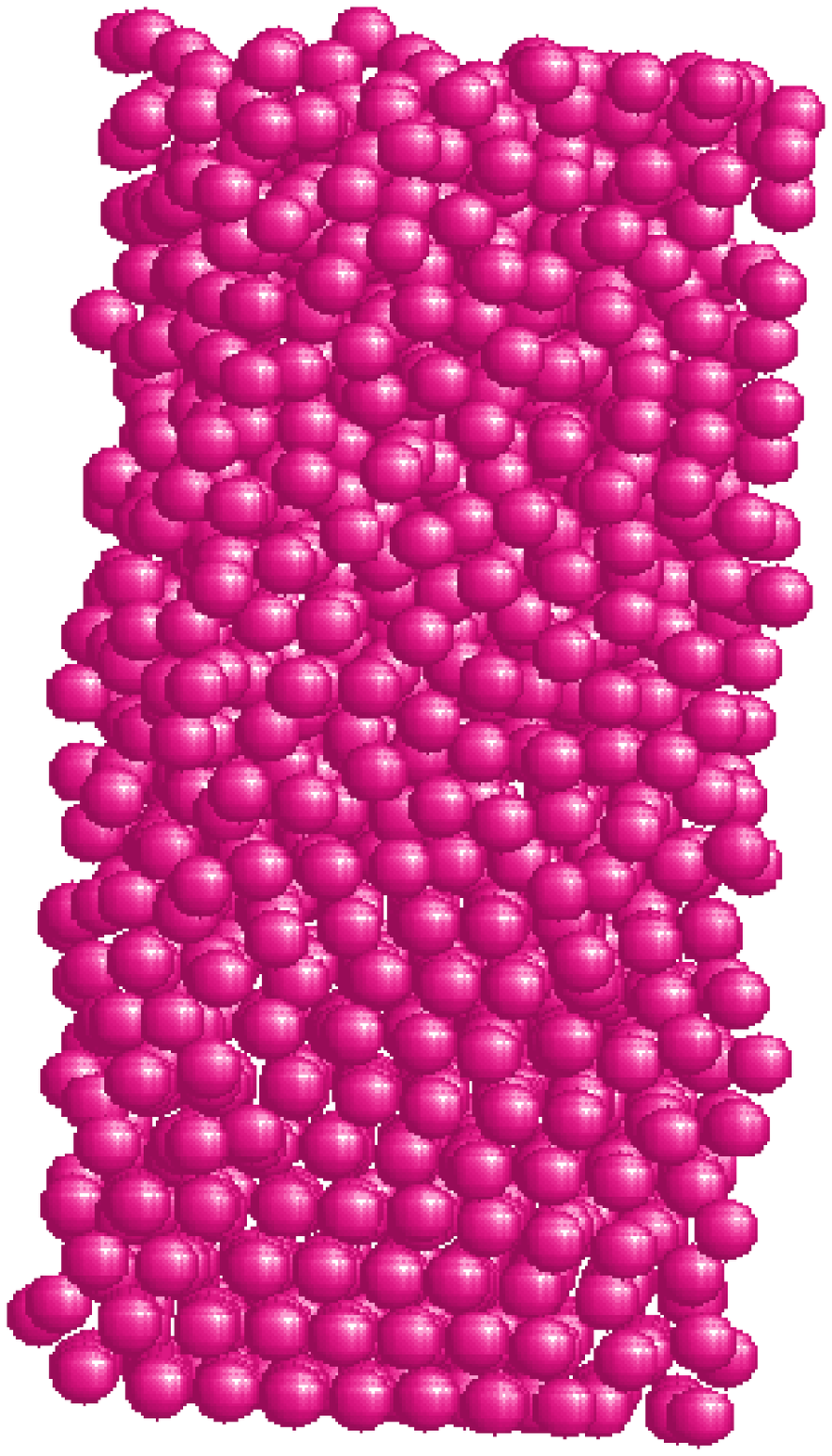}} \epsfxsize=2.5truecm\epsfysize=4truecm
{\epsfbox{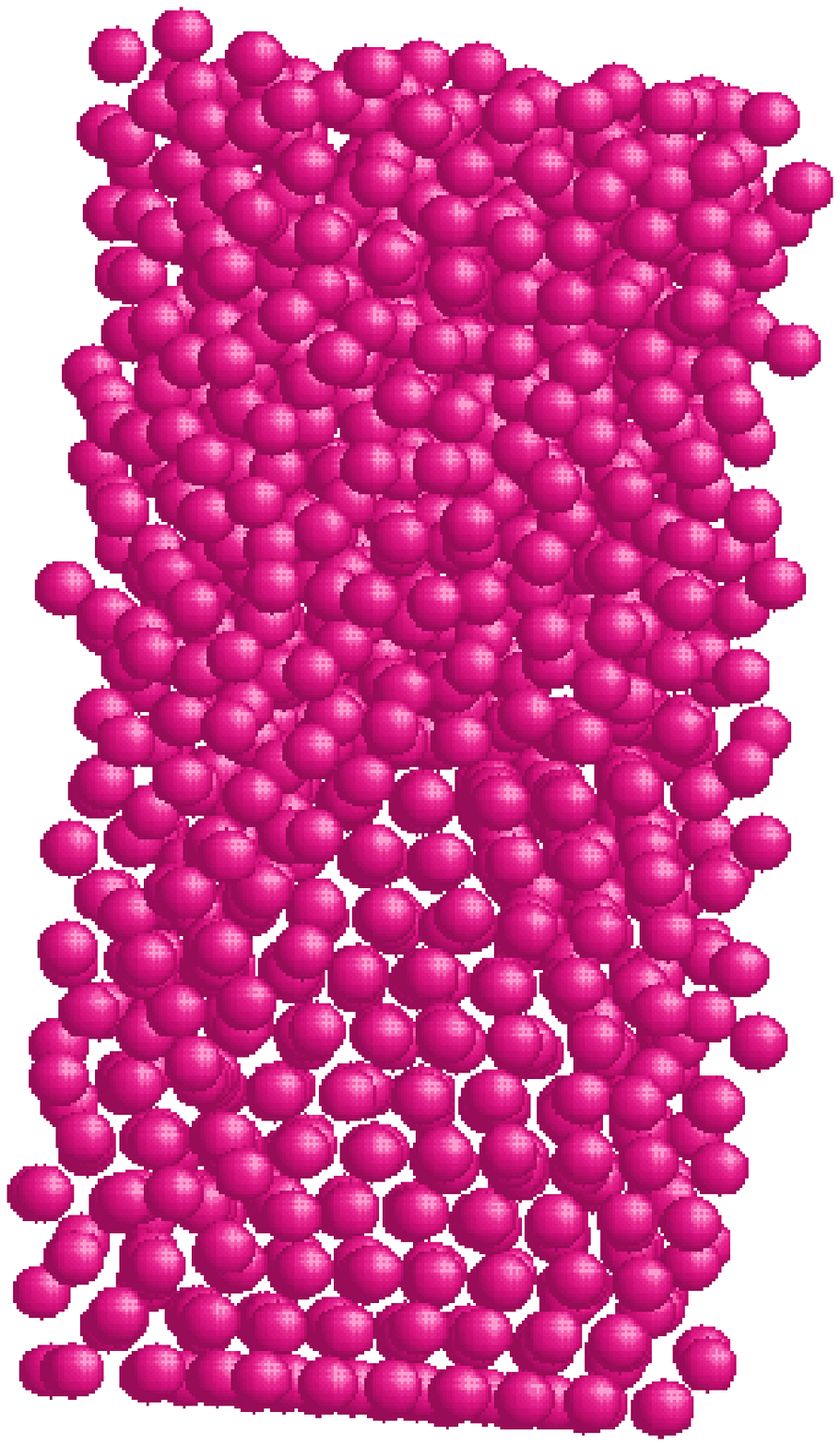}} \epsfxsize=2.5truecm\epsfysize=4truecm
{\epsfbox{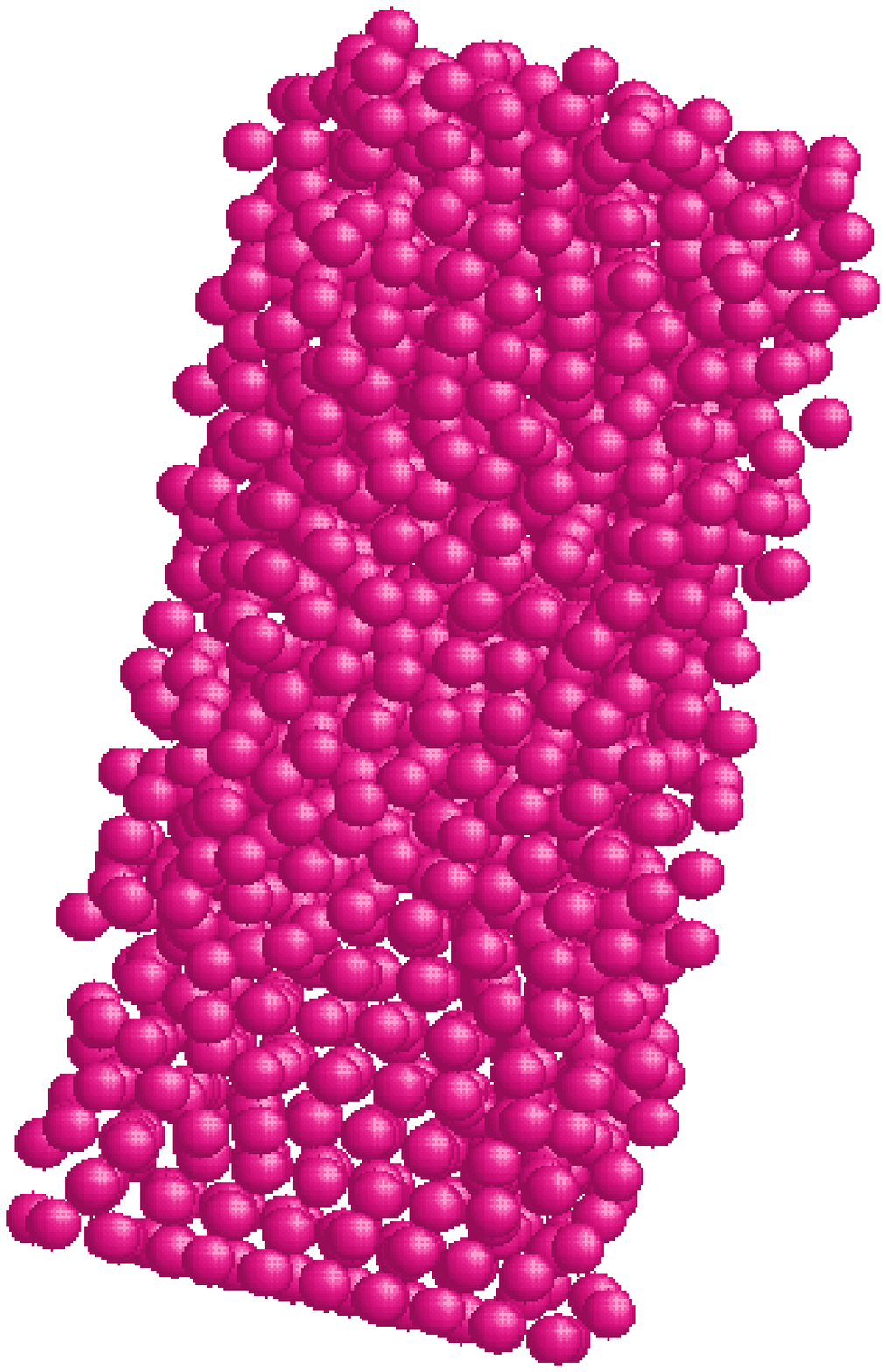}} \epsfxsize=2.5truecm\epsfysize=4truecm
{\epsfbox{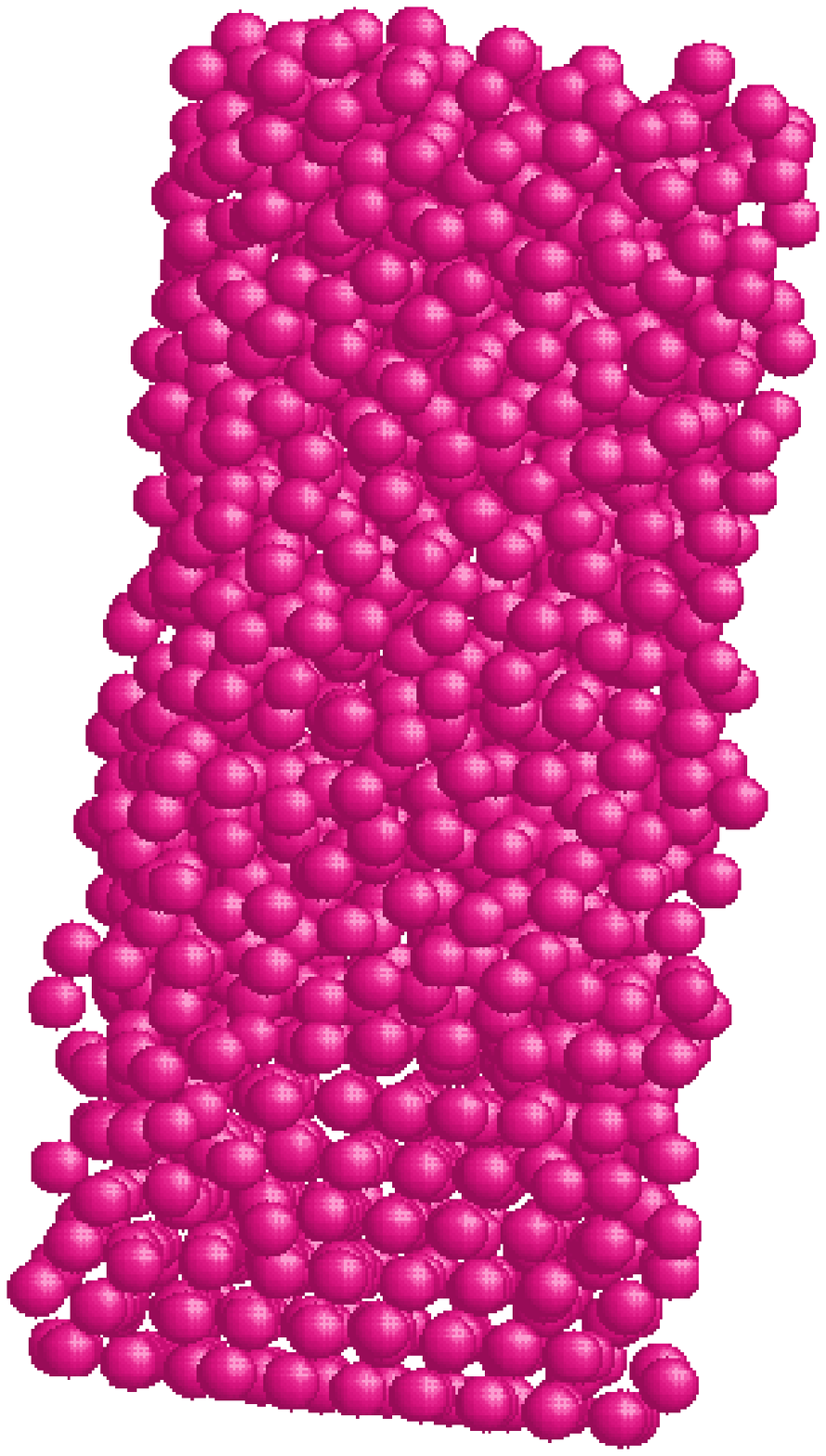}} \epsfxsize=2.5truecm\epsfysize=4truecm
{\epsfbox{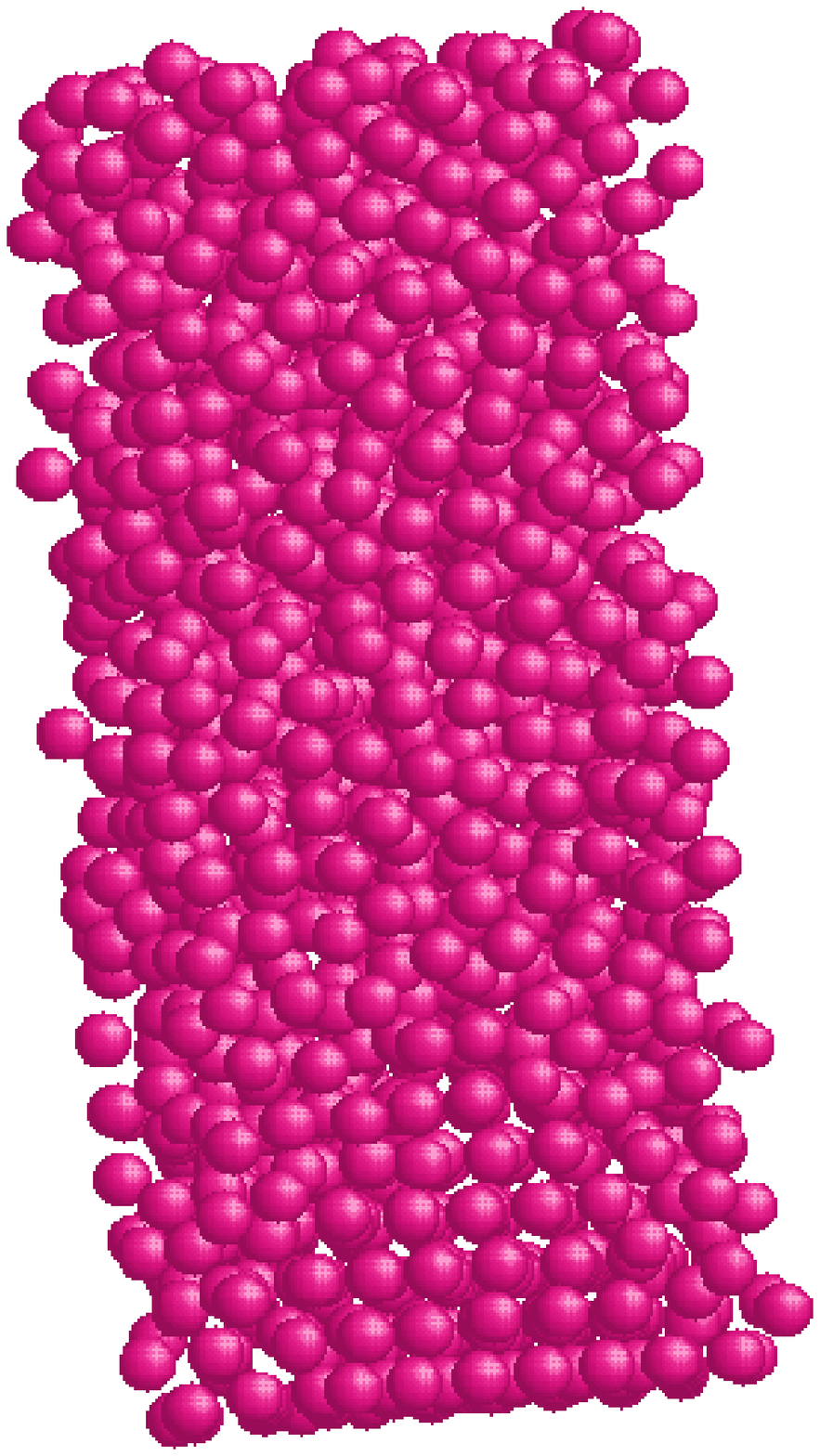}} 
\newline
\caption{ Configurations for selected evenly spaced shaking
amplitudes $A$, a) on  the irreversible branch $A=0.3, 0.6, 0.9$,
b) the `reversible' branch for $A$ decreasing $A=0.8, 0.5, 0.2$,
and finally c) the `reversible' branch for $A$ increasing $A=
0.15, 0.45, 0.75$}.
\end{center}
\end{figure}

Investigations are under way to analyse these structures further,
with the use of appropriate order parameters \cite{surajit} in
the context of an observed crystallisation transition in shaken
granular beds \cite{jpcm}.

\section{Discussion}
We have presented a simple model of short time inhomogeneous
consolidation, based on the idea of a consolidation wave; its
analytical solution is in good agreement with our Monte Carlo
simulations of vibrated (three-dimensional) particle packings.
Despite the simplicity of our model, it manifests an important
signature of inhomogeneous consolidation for early and
intermediate times, in agreement with the simulations.
Interestingly these are both in agreement with experimental data
\cite{knight}, where it was shown that the larger the intensity
of vibration, the more homogeneous the consolidation ; notice
that the data for the larger vibrational intensity asymptote much
more quickly (Figure \ref{Fig3}). Investigations are in progress
to look in greater detail at the local structures in the particle
packings, and at the detailed spatial variations inherent in the
inhomogeneous packings \cite{surajit}. Ongoing work also includes a detailed
investigation of inhomogeneous density fluctuations in a
vibrated powder, and their repercussions for its power spectrum
\cite{ednowak}.

Additionally, we have presented ancillary evidence for
consolidation waves via the study of particulate
configurations in the annealed shaking of a granular bed. In addition
to the aspects on which we have reported here,
an analytical study of a simple lattice-based model \cite{peter} of a shaken
granular bed is expected to yield valuable insights on the
precise influence of spatial inhomogeneities along the reversible
and irreversible branches of the compaction curves.

\section{Acknowledgement}
GCB acknowledges support from the Biotechnology and Biological
Sciences Research Council, UK (218/FO6552).

\end{document}